\documentclass[aps,prl,twocolumn,showpacs,superscriptaddress,groupedaddress,longbibliography]{revtex4-2}  
\usepackage{graphicx,comment,float,braket,textcomp,amsmath,lipsum}  
\usepackage{dcolumn}   
\usepackage{bm}        
\usepackage{amssymb}   

\begin{document}

\title{Spectral Broadening and Ultrafast Dynamics of a Nitrogen-Vacancy Center Ensemble in Diamond}

\author{Albert Liu}
\thanks{Present address: Max Planck Institute for the Structure and Dynamics of Matter, Hamburg, Germany}
\affiliation{Department of Physics, University of Michigan, Ann Arbor, Michigan 48109, USA}

\author{Diogo B. Almeida}
\affiliation{Department of Physics, University of Michigan, Ann Arbor, Michigan 48109, USA}
\affiliation{Instituto de Fisica "Gleb Wataghin", Universidade Estadual de Campinas, 13083-859, Campinas, Sao Paulo, Brazil}

\author{Ronald Ulbricht}
\affiliation{Max Planck Institute for Polymer Research, Ackermannweg 10, 55128 Mainz, Germany}

\author{Steven T. Cundiff}
\email{cundiff@umich.edu}
\affiliation{Department of Physics, University of Michigan, Ann Arbor, Michigan 48109, USA}

\vskip 0.25cm

\date{\today}

\begin{abstract}
    Many applications of nitrogen-vacancy (NV) centers in diamond crucially rely on a spectrally narrow and stable optical zero-phonon line transition. Though many impressive proof-of-principle experiments have been demonstrated, much work remains in engineering NV centers with spectral properties that are sufficiently robust for practical implementation. To elucidate the mechanisms underlying their interactions with the environment, we apply multi-dimensional coherent spectroscopy to an NV center ensemble in bulk diamond at cryogenic temperatures. Our spectra reveal thermal dephasing due to quasi-localized vibrational modes as well as ultrafast spectral diffusion on the picosecond timescale. The intrinsic, ensemble-averaged homogeneous linewidth is found to be in the tens of GHz range by extrapolating to zero temperature. We also observe a temperature-dependent Stark splitting of the excited state manifold, relevant to NV sensing protocols.
\end{abstract}

\pacs{}
\maketitle

Vacancy centers in diamond \cite{Aharonovich2014,Schroder2016} have attracted sustained interest due to their exceptional optical properties. In particular, nitrogen-vacancy (NV) centers \cite{Doherty2013}, the most well-studied of all defects in diamond, has found a variety of potential applications such as nanoscale sensors \cite{Schirhagl2014}, solid-state qubits \cite{Dobrovitski2013}, and single-photon emitters \cite{Aharonovich2016}.

However, an obstacle to practical implementation of NV centers in devices is their non-zero electric dipole moment that leads to spectral instability and linewidth broadening \cite{Acosta2012}. Other defects with inversion symmetry, most notably silicon-vacancy (SiV) centers, have proven less sensitive to their environment but possess their own disadvantages such as short spin lifetimes and low internal quantum efficiencies (i.e. up to around 10\% for SiV centers \cite{Neu2012} compared to over 70\% for NV centers \cite{Radko2016}). There has thus been much recent effort towards engineering environmentally insensitive NV centers \cite{Chu2014,vanDam2018}, but doing so rationally will require a greater understanding of the fundamental interactions between NV centers and their environment.

Specifically, the dynamics of NV centers at both ultrafast timescales ($\leq$ ps) and low temperatures remains largely unexplored. In this regime, physical processes that lead to environmental perturbations of NV centers may be characterized by observing broadening of their optical transition linewidth \cite{Liu2020}. We further desire their {\it ensemble-averaged} properties, which inform us about the underlying physics without being biased by single exceptional centers. Unfortunately, the homogeneous properties of NV centers are obscured in ensemble measurements by inhomogeneous broadening that arises from locally varying electric field and strain.

We circumvent the inhomogeneous broadening of an NV center ensemble in bulk diamond by applying four-wave mixing (FWM) spectroscopy at cryogenic temperatures. First, multi-dimensional coherent spectroscopy (MDCS) is performed to extract the maximum amount of information from the system's nonlinear optical response. The measured third-order optical response exhibits a broad inhomogeneous lineshape, twice the inhomogeneous width of linear ensemble spectra, comprised of two distinct orbital state distributions which exhibit dramatic shifts with changing temperature. Temperature- and waiting time-dependent measurements then determine the dominant linewidth-broadening mechanisms to be quasi-localized phonon-induced dephasing and spectral diffusion on ultrafast timescales. Complementary time-domain integrated FWM measurements are performed, which reveal non-Markovian dephasing dynamics on the picosecond timescale. These results provide crucial insight into the decoherence processes in NV centers that limit quantum technologies. 

\section{Materials and Methods}

\begin{figure}[t]
    \centering
    \includegraphics[width=0.48\textwidth]{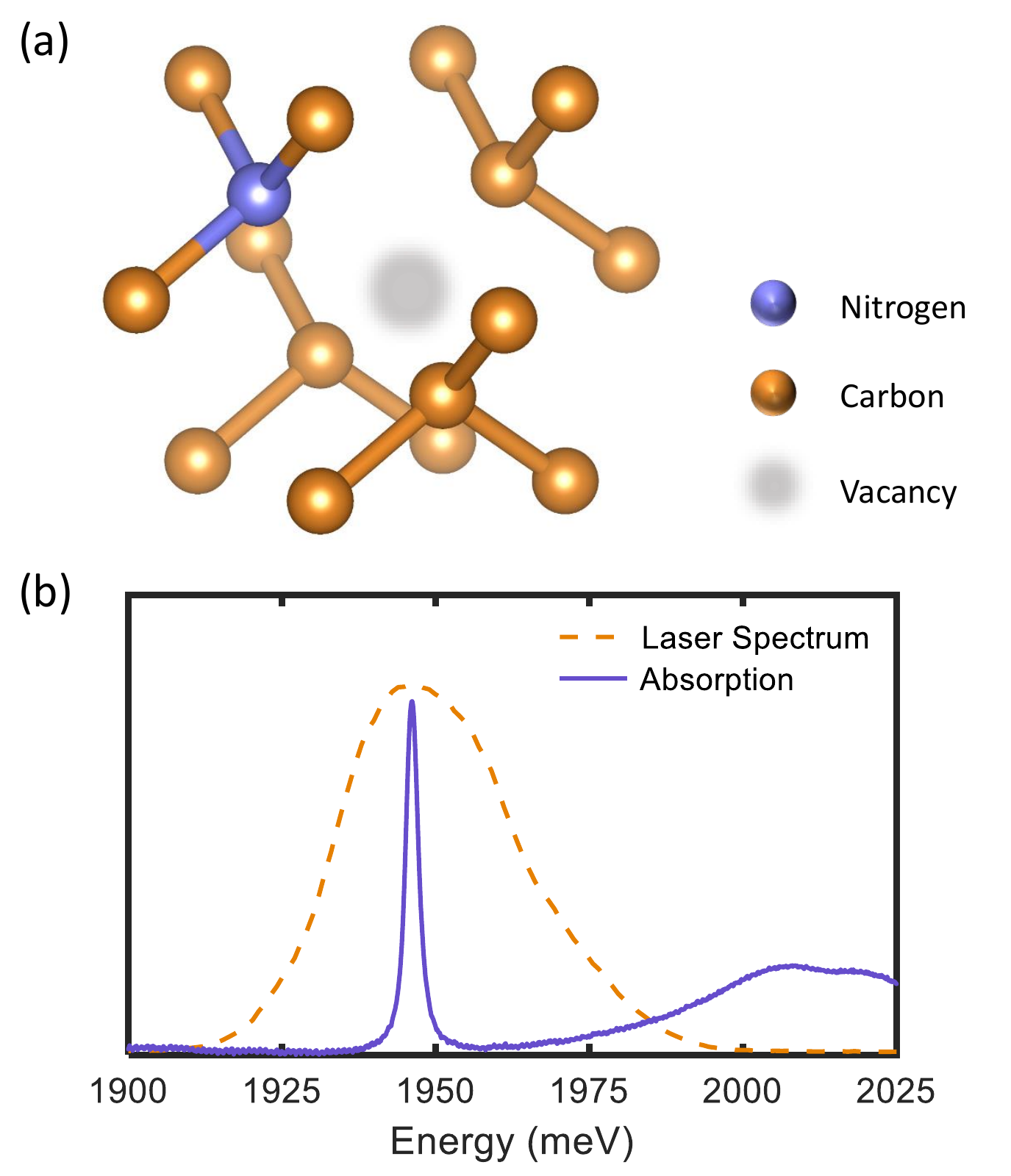}
    \caption{(a) Diagram of a nitrogen vacancy center in diamond. (b) Absorption spectrum (purple) of the studied sample at 20 K. A narrow zero-phonon line is observed at 1946 meV (637 nm) next to a broad phonon sideband that extends to higher energies. The spectrum of the MDCS excitation pulses is also shown by the dashed orange line. Both curves are normalized for comparison.}
    \label{Fig1}
\end{figure}

The sample studied is type Ib bulk monocrystalline diamond, where vacancy centers were introduced by irradiation with 1 MeV electrons and subsequent annealing, resulting in an NV$^-$ density of 1-2~ppm. The atomic structure of such a center is shown in Fig.~\ref{Fig1}a, which consists of a vacancy point-defect and an adjacent nitrogen substitution. This configuration gives rise to a cryogenic absorption spectrum consisting of a narrow zero-phonon line (ZPL) and an adjacent phonon sideband, as shown in Fig.~\ref{Fig1}b. We center our excitation laser spectrum on the ZPL, and our narrow pulse bandwidth (compared to the known discrete vibrational modes \cite{Huxter2013,Ulbricht2018-1}) means that we predominantly measure the ZPL resonance of NV$^{-}$ without other phonon-assisted transitions. The ZPL center energy is known to shift with temperature \cite{Davies1974,Doherty2014} due to lattice interactions \cite{ThreeOscillatorPaper}, but is relatively stationary in our measurements performed below 150 K. NV centers in their neutral charge state (NV$^0$) have a higher energy transition energy of 2156~meV (575~nm) and are not excited in our experiment.

\begin{figure}[b]
    \centering
    \includegraphics[width=0.5\textwidth]{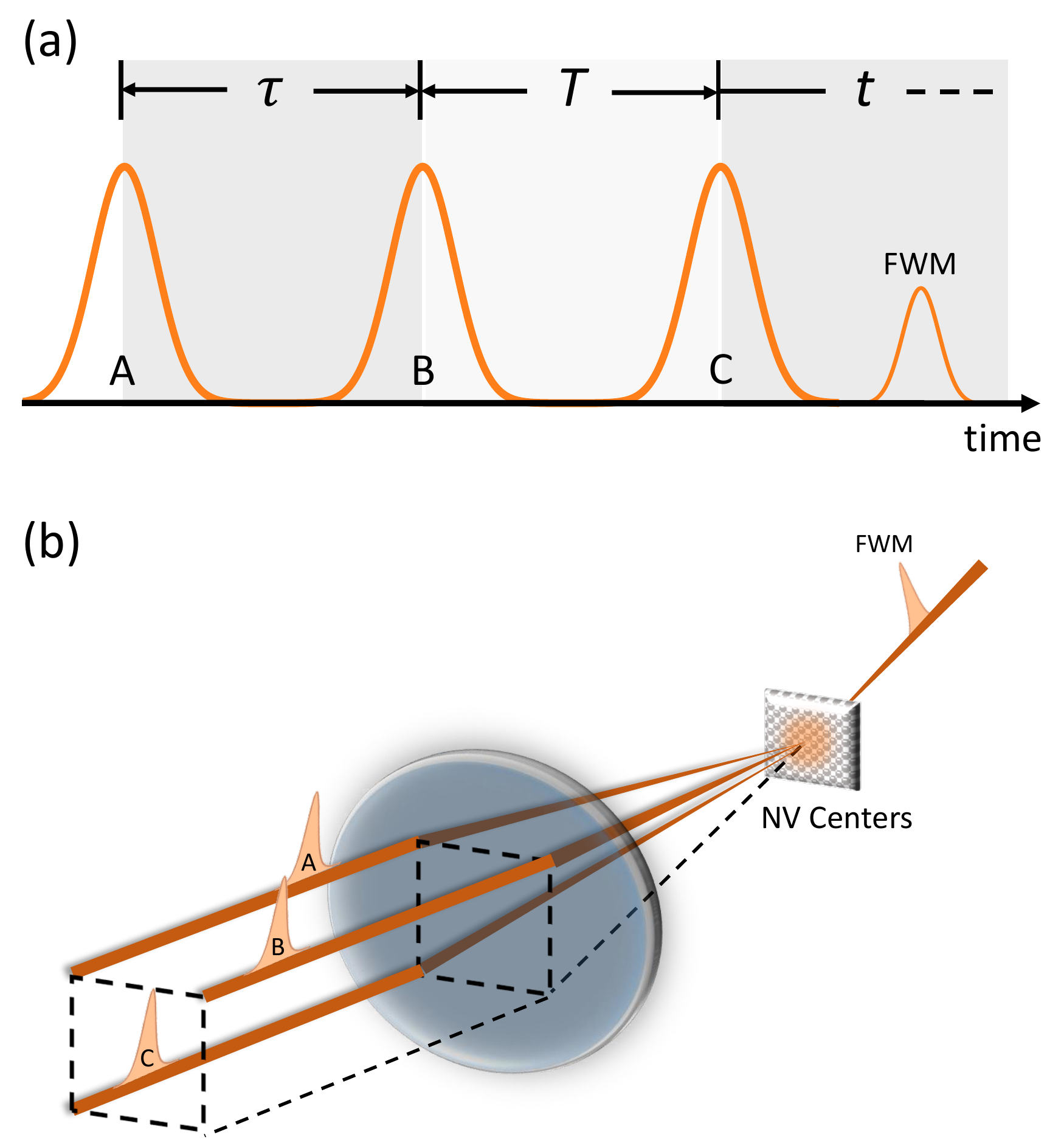}
    \caption{(a) Diagram defining the three time-delays $\tau$, $T$, and $t$ that we Fourier transform along to generate an MDCS spectrum. (b) Experimental schematic, consisting of three pulses arranged in the box geometry that generate a photon echo.}
    \label{Fig2}
\end{figure}

Three resonant laser pulses are used to generate a FWM signal as shown in Fig.~\ref{Fig2}a. To perform MDCS, the FWM signal is recorded as a function of three time-delays and subsequent Fourier transforms along two or all three variables returns a multi-dimensional spectrum. Here we Fourier transform an emitted photon echo along delays $\tau$ and $t$, which returns a so-called one-quantum spectrum \cite{Liu2019-1,Liu2019-2}. By correlating absorption and emission dynamics, a one-quantum spectrum unambiguously separates homogeneous and inhomogeneous broadening in orthogonal directions \cite{Siemens2010,Liu2019-1,Liu2019-2}.

A schematic of our MDCS experiment is shown in Fig.~\ref{Fig2}b, in which three pulses are focused onto our sample in the box geometry to generate a photon echo FWM signal that is heterodyne detected with a separate local-oscillator pulse routed around the sample. The pulses used are approximately 90~fs in duration and generated by an optical parametric amplifier at a repetition rate of 250~kHz. The laser excitation density used to obtain the data is 1~W/cm$^2$ except for that in Fig.~\ref{Fig4} where it is 3.3~W/cm$^2$ (both verified to generate predominately third-order responses via power-dependence measurements). The sample is mounted to a cold-finger cryostat and cooled to cryogenic temperatures.

\begin{figure*}
    \centering
    \includegraphics[width=0.99\textwidth]{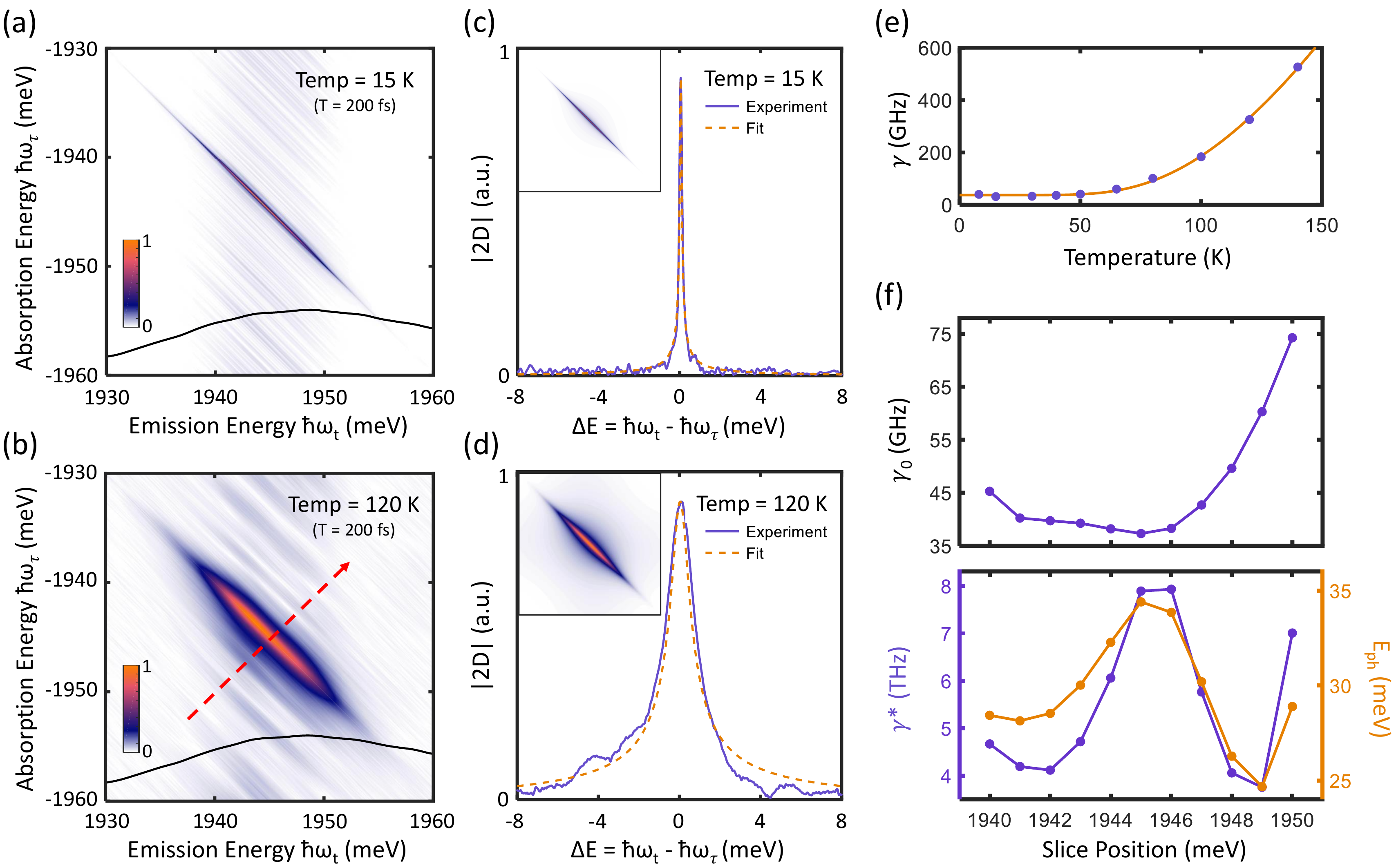}
    \caption{One-quantum spectra taken at $T$ = 200~fs and temperatures (a) 15~K and (b) 120~K. The solid black line indicates the respective excitation laser spectrum at each temperature. Elongation along the diagonal direction reflects inhomogeneous broadening of the ZPL transition. (c-d) Experimental and fitted cross-diagonal lineshapes are plotted at each temperature, with the resultant one-quantum spectra inset. The slice positions at $|\hbar\omega_\tau| = |\hbar\omega_t| = 1945$ meV are indicated by the dashed red arrow in Fig.~\ref{Fig3}b. (e) Fitted values of $\gamma$ at temperatures increasing from 6~K to 140~K. The exponential increase in $\gamma$ is fitted to a localized phonon dephasing model and plotted as the solid orange curve. (f) Homogeneous thermal dephasing parameters as a function of resonance energy. Values are obtained by repeating the fitting procedure in (e) for slices taken at $|\hbar\omega_\tau| = |\hbar\omega_t| =$ 1940 meV to 1950 meV.}
    \label{Fig3}
\end{figure*}

\section{Results and Discussion}

\subsection{Multi-Dimensional Coherent Spectroscopy}

We examine the dephasing dynamics by resolving homogeneous lineshapes in one-quantum spectra. One-quantum spectra at temperatures 15~K and 120~K are shown in Fig.~\ref{Fig3}a and \ref{Fig3}b, in which the vertical $\hbar\omega_\tau$ axes are opposite in sign to the horizontal $\hbar\omega_t$ axes due to inverse phase-evolution during delays $\tau$ and $t$. Strong inhomogeneous broadening of the ZPL transition manifests as peak elongation along the diagonal ($|\hbar\omega_\tau| = |\hbar\omega_t|$) direction \cite{Siemens2010}, while lineshapes in the orthogonal cross-diagonal direction reflect homogeneous broadening. However, homogeneous ($\gamma = 1/T_2$) and inhomogeneous (standard deviation $\sigma$ of a Gaussian transition frequency distribution) broadening are completely separated in each direction only for systems dominated by inhomogeneous broadening ($\sigma \gg \gamma$). To extract $\gamma$ and $\sigma$ in the presence of moderate inhomogeneous broadening, as in NV centers, lineshapes in the diagonal and cross-diagonal directions must be simultaneously fit \cite{Siemens2010}. The experimental and fitted cross-diagonal lineshapes for each temperature are shown in Fig.~\ref{Fig3}c and \ref{Fig3}d, with the resultant one-quantum spectra plotted inset.

A temperature dependence of $\gamma$ measured for a resonance energy of 1945 meV is plotted in Fig.~\ref{Fig3}e, where the observed exponential increase in dephasing rate $\gamma$ with temperature is characteristic of pure dephasing due to elastic interactions with discrete phonons. The thermal dephasing may be modeled by a linear dependence on the phonon mode occupation \cite{Singh2013}:
\begin{align}
    \gamma(T_s) = \gamma_0 + \frac{\gamma^*}{e^{E_{\text{ph}}/k_BT_s} - 1}
\end{align}
where $\gamma_0$ is the intrinsic zero-temperature dephasing rate, $\gamma^*$ is the thermal dephasing amplitude, $T_s$ is the sample temperature, and $E_{ph}$ is the energy of the involved phonon mode. A least-squares fit to our data is plotted as a solid line in Fig.~\ref{Fig3}e, with fitted parameters $\gamma_0 = 37.31$~GHz ($T_2 = 1/\gamma_0 = 26.8$ ps), $\gamma^* = 7890$~GHz, and $E_{\text{ph}} = 34.41$~meV. The phonon energy $E_{ph}$ is in very good agreement with previous experimental results on the activation energy for ZPL dephasing from single-defect linewidth measurements \cite{Fu2009} and transient absorption spectroscopy \cite{Ulbricht2016}. This phonon mode was calculated to be a vibronic $A_1$ mode originating from the Jahn-Teller distortion in the electronic, doubly-degenerate excited state \cite{Abtew2011}. Another simulation found a quasi-localized vibrational mode of similar energy that involves vibrations of carbon atoms almost exclusively \cite{Zhang2011}.

We also repeat our linewidth temperature dependence analysis for cross-diagonal lineshapes taken at positions ranging from $|\hbar\omega_\tau| = |\hbar\omega_t| = 1940$ meV to 1950 meV, corresponding to varying resonance energy. The slice position dependence of the three thermal broadening parameters $\gamma_0$, $\gamma^*$, and $E_{ph}$ are plotted in Fig.~\ref{Fig3}f. We find that the zero-temperature linewidth $\gamma_0$ sharply increase for energies higher than 1946 meV, and that both $\gamma^*$ and $E_{ph}$ vary strongly throughout the inhomogeneous distribution. These observations provide crucial information concerning the underlying mechanism of inhomogeneous broadening in this system.

We note that an activation energy of $E_{\text{ph}} \approx 4.84$ meV was previously reported by integrated FWM measurements \cite{Lenef1996} which, measured at a long waiting time of $T = 1.5$~ns, is inconsistent with our results likely due to spectral diffusion. Femtosecond polarization anisotropy measurements in \cite{Ulbricht2016}, which probe the depolarization of the excitation-induced coherence between the two excited state orbitals, reported a decoherence time at the low temperature limit of $T_1$ = 14.4 ps. This perfectly matches the $T_2$ values reported here. Note that the significant spectral bandwidth of the femtosecond pulses used in both studies necessarily excites both orbital states. This differs from cryogenic PLE experiments on single NV centers \cite{Shen2008,Fu2009}, in which an optical coherence is induced between the ground state and only a single orbital by means of narrowband laser excitation. The discrepancy between time-resolved dephasing lifetime measurements (here and elsewhere \cite{Lenef1996}) and PLE measurements that report radiatively-limited optical coherence times may thus originate from modification of orbital relaxation due to simultaneous excitation of the entire excited state manifold.

\begin{figure}[t]
    \centering
    \includegraphics[width=0.5\textwidth]{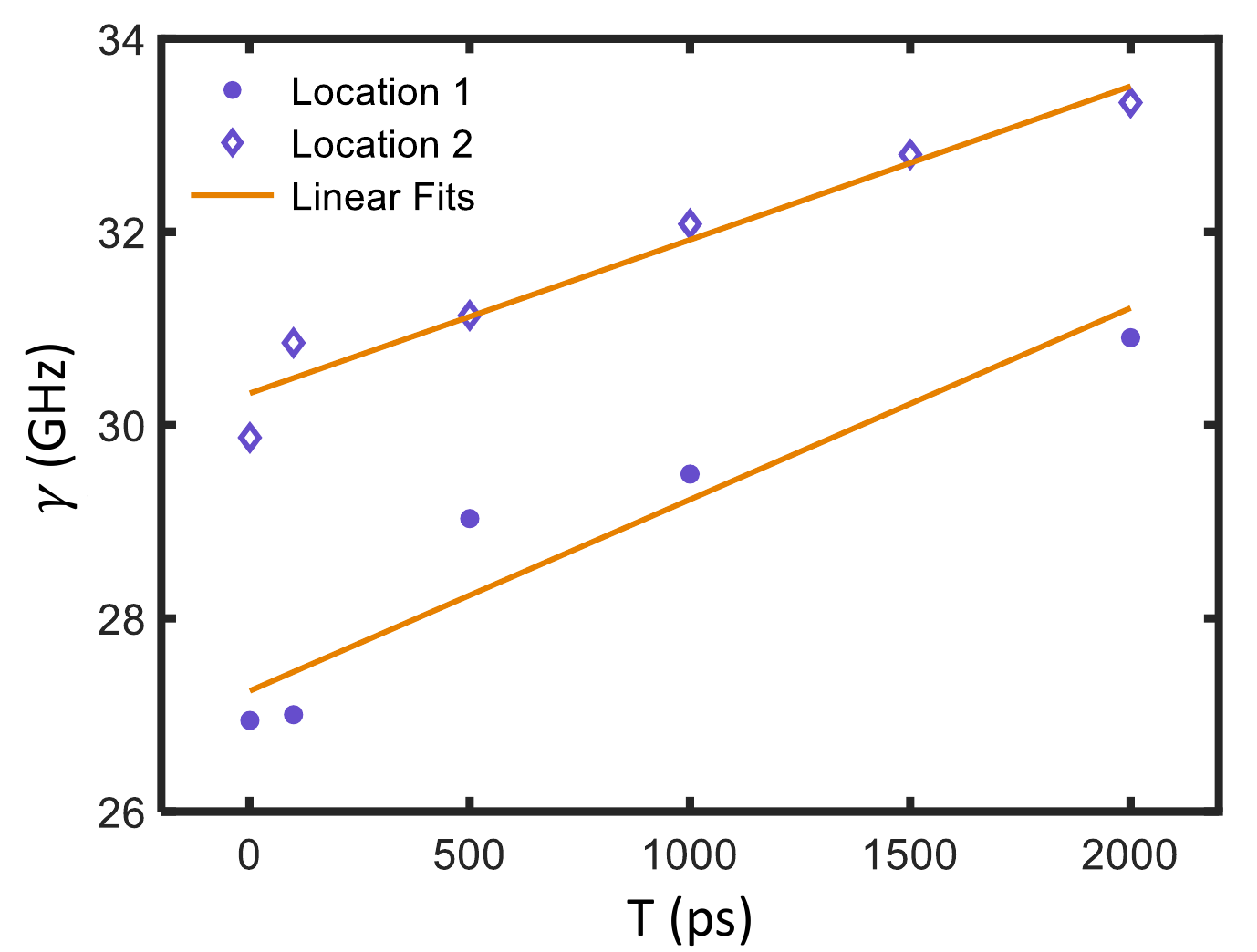}
    \caption{Fitted values of $\gamma$, taken at two different sample locations and 10~K, as a function of waiting time $T$. A monotonic increase in linewidth is observed as $T$ increases from 1~ps to 2~ns, which is fitted as shown to linear spectral diffusion rates of 1.98~MHz/ps and 1.59~MHz/ps at each location.}
    \label{Fig4}
\end{figure}

\subsection{Ultrafast Spectral Diffusion}

MDCS also enables the powerful capability to probe ultrafast spectral diffusion. While pulses $A$ and $C$ probe absorption and emission dynamics of the sample, the second pulse may be thought to induce a population state whose energy will vary in time. As the waiting time $T$ between pulses $B$ and $C$ increases, absorption and emission energies of a resonance become less correlated and broadening in the cross-diagonal direction occurs. For systems with strong spectral diffusion, simultaneous broadening and distortion of the cross-diagonal lineshape occurs, which allows extraction of the frequency-frequency correlation function \cite{Singh2016} that quantifies the resonance energy fluctuations. However, we find that broadening of NV center lineshapes is too weak (within our experimental $T$ delay limit) to perform such an analysis. We therefore plot a dependence of $\gamma$ on waiting time $T$ in Fig.~\ref{Fig4}, which provides an effective measure of spectral diffusion at ultrafast timescales. From measurements taken at two different locations on our sample, as $T$ is increased from 1 ps to 2 ns the fitted effective dephasing rate $\gamma$ increases at 1.98~MHz/ps and 1.59~MHz/ps. In a previous photon correlation study of NV centers in nanodiamonds \cite{Wolters2013} ultrafast spectral diffusion was found to occur only with simultaneous 532 nm excitation, which implicated impurity photo-ionization and subsequent charge trapping as the underlying mechanism. Our measurements of spectral diffusion, which involve only resonant excitation of the ZPL, indicate that spectral diffusion of NV centers in bulk diamond may occur due to other mechanisms such as reorganization of the surrounding diamond lattice following optical excitation.

\begin{figure}[t]
    \centering
    \includegraphics[width=0.5\textwidth]{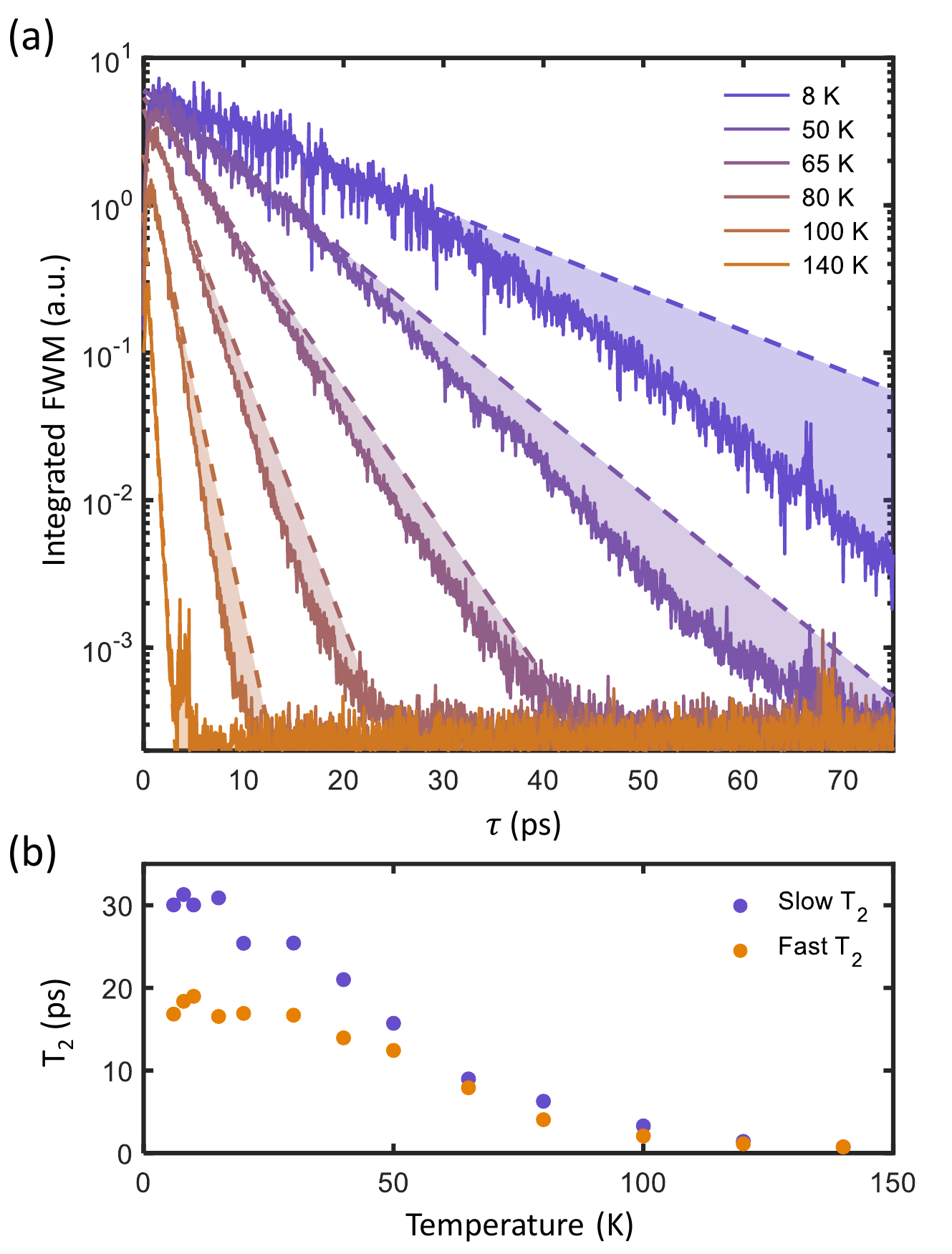}
    \caption{(a) Integrated FWM field taken at $T = 200$ fs as a function of delay $\tau$. Dashed lines represent exponential fit to initial, slower dephasing. Shaded areas indicates deviation from the initial dephasing rate as decoherence accelerates. (b) Values of $T_2$ fitted from earlier (purple) and later (orange) ranges of $\tau$.}
    \label{Fig5}
\end{figure}

\subsection{Integrated Four-Wave Mixing}

The ultrafast spectral diffusion observed comprises a microscopic mechanism of pure dephasing \cite{Mukamel1999}. For excitations in crystals (such as the case for NV centers), resonance energy fluctuations primarily arise from frequency-dependent interactions with their thermal environment, which often gives rise to memory-effects in decoherence \cite{Carmele2019}. While these memory-effects are difficult to observe in the frequency-domain, they clearly manifest in the time-domain as non-exponential (non-Markovian) dephasing \cite{Lorenz2005,Liu2019-1}.

We therefore measure the integrated photon-echo field as a function of delay $\tau$. In the case of memoryless (Markovian) dephasing, the photon-echo field decays exponentially in $\tau$ with a time-constant identical to twice the dephasing time $2T_2$ of an inhomogeneously-broadened optical transition \cite{Lorenz2005}. We note however, that this dephasing time is an effective value averaged over the inhomogeneous resonance frequency distribution. Measurements performed at representative temperatures are shown in Fig.~\ref{Fig5}a, which exhibit an increase in dephasing rate with temperature. Unexpectedly, the dephasing dynamics exhibit two distinct time constants with more rapid dephasing occuring at longer $\tau$. Such behavior is reminiscent of cross-over between Markovian and non-Markovian dephasing, previously observed in atomic vapors \cite{Lorenz2005}. Separate fitting of each segment yields two dephasing times, shown in Fig.~\ref{Fig5}b, that differ by over 10 ps at the lowest temperatures but converge as temperature increases. The longer fitted dephasing times in Fig.~\ref{Fig5}b at low temperatures agree well with the intrinsic dephasing rates extracted from our 2D measurements shown in Fig.~\ref{Fig3}. Interestingly, the values and temperature dependence of the fast decay component perfectly match the dephasing times determined through previous transient polarization anisotropy measurements \cite{Ulbricht2016}.

\begin{figure}[b]
    \centering
    \includegraphics[width=0.5\textwidth]{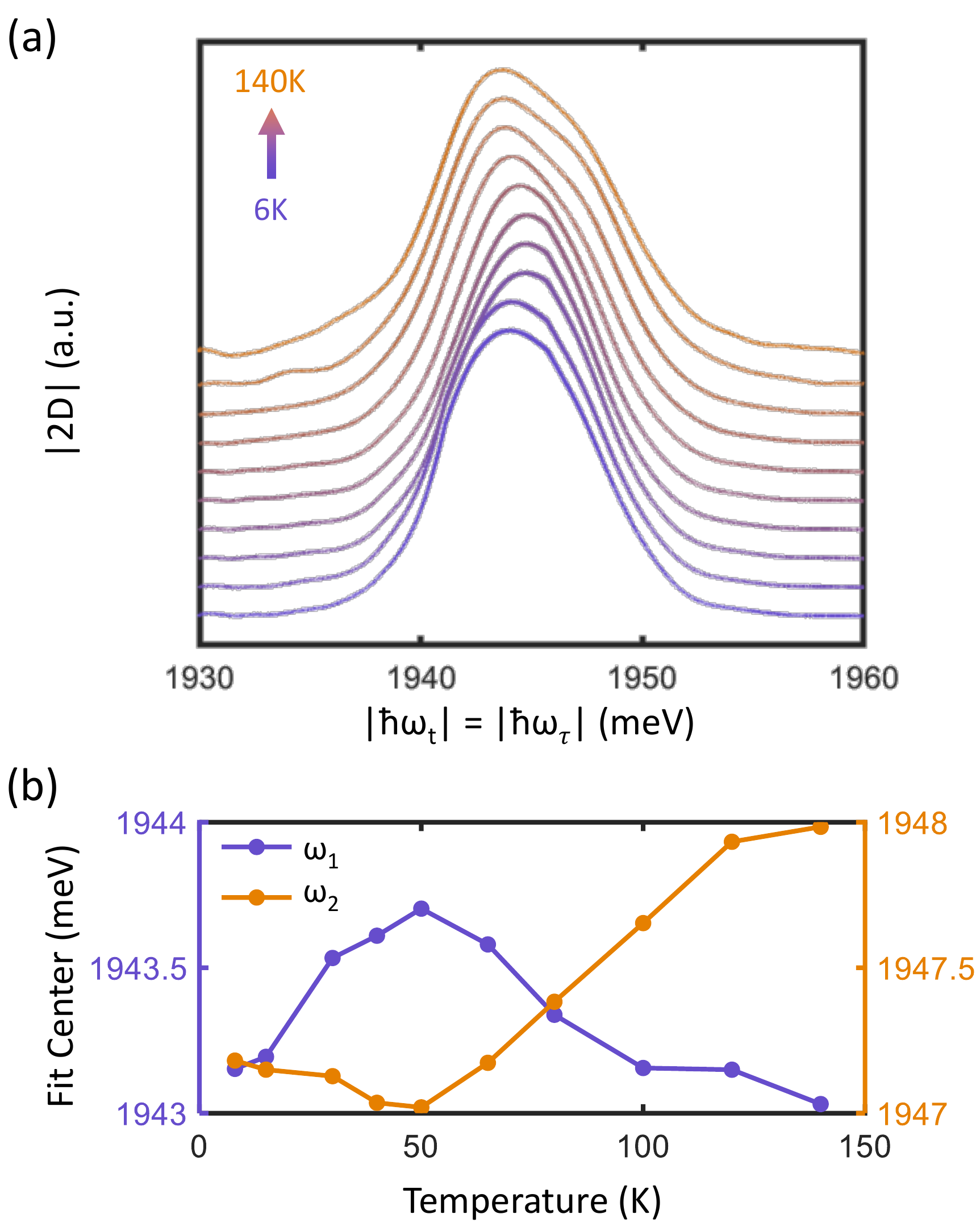}
    \caption{(a) Respective diagonal slices at the same temperatures in Fig.~\ref{Fig3}(e). (b) Fitted center energies $\omega_1$ and $\omega_2$ of two Gaussian distributions used to model the diagonal slices in (a). Fits are plotted in the Supplemental Info.}
    \label{Fig6}
\end{figure}

\subsection{Bimodal Inhomogeneous Distribution}

We next examine diagonal slices of the 2-D spectra taken along $|\hbar\omega_\tau| = |\hbar\omega_t|$ (plotted in Fig.~\ref{Fig6}a as a function of temperature). Interestingly, increasing temperature results in the formation of two distinct lobes in the diagonal slice lineshape, not visible in the linear absorption spectra (see Supplemental Info), which we attribute to electric field-induced splitting \cite{Tamarat2006,Bassett2011} of the $^3E$ manifold. An absence of sidebands in the one-quantum spectra also indicates that the two families observed are not coherently coupled. We thus modeled this system as two independent Gaussian distributions, each corresponding to an excited orbital state. The two Gaussians are of fixed widths $\sigma_1 = 2.6$ meV (627.8 GHz) and $\sigma_2 = 2.3$ meV (555.4 GHz), with only center frequencies allowed to vary, which reproduced the data well (see Supplemental Info). The fitted center frequencies as a function of temperature are plotted in Fig.~\ref{Fig6}b, which vary in the single meV range and appear anti-correlated. Indeed, symmetric splitting of the excited orbital states has recently been predicted and observed in NV center ensembles in response to perpendicularly-oriented electric fields \cite{Block2020}. Remarkably, our observed splitting of $\approx$ 5 meV (40 cm$^{-1}$) perfectly matches the values reported from photon echo \cite{Lenef1996} and ODMR \cite{vanOort1991} experiments, where the former report \cite{Lenef1996} proposed a Jahn-Teller splitting in the excited state manifold that has since been disputed \cite{Goss1997}. According to the transverse electric-field susceptibility of $\chi^e_\perp = 1.4$ MHz/(V/cm) reported by Block et al. \cite{Block2020}, the splittings we observe correspond to fields ranging from 0.29 MV/cm at 50 K to 0.43 MV/cm at 140 K. These internal fields are possibly generated by positively charged single substitutional nitrogen defects in close proximity to NV centers \cite{Manson2018}.

\section{Conclusion}

In conclusion, we have measured both integrated FWM and one-quantum spectra of NV centers in diamond, which simultaneously provide both the homogeneous dephasing rate and inhomogeneous resonance frequency distribution. The primary ZPL dephasing mechanism is found to be a Jahn-Teller-induced vibronic state, in accordance with previous studies. In addition, waiting time-dependent measurements reveal another source of extrinsic broadening due to ultrafast spectral diffusion that occurs solely from resonant photo-excitation. Knowledge of these broadening mechanisms provide crucial, fundamental knowledge for engineering spectrally stable NV centers with narrow optical linewidths, with applications toward quantum information protocols \cite{Sipahigil2012,Bernien2013}. Lastly we observe Stark-splitting of excited orbital states that is not visible in linear spectroscopy and forms a bimodal structure in the inhomogeneous optical response, and whose center frequencies show distinct temperature dependencies. With recent advances in the rapid acquisition of multi-dimensional spectra \cite{Lomsadze2017,Lomsadze2018}, our results suggest a microwave-free, all-optical analogue of a recently proposed electric-field sensing protocol \cite{Block2020} in NV center ensembles.

We acknowledge N. Manson for fruitful discussions. This work was supported by the Department of Energy grant number DE-SC0015782. D.B.A. acknowledges support by a fellowship from the Brazilian National Council for Scientific and Technological Development (CNPq). R.U. acknowledges support by a DFG fellowship (UL 474/1-1).

\bibliography{bibliography}

\end{document}